%% file: ArxivVC_5.tex
\documentclass[10pt,a4paper,fleqn]{article}
\usepackage{amsmath}
\usepackage{dingbat}
\usepackage{dingbat}
\usepackage{amssymb}
\usepackage{fancybox}
\usepackage{amsthm}
\usepackage{fancyhdr}
\usepackage{pstricks}
\usepackage{pst-plot}
\usepackage{graphicx}
\usepackage{ulem}
\usepackage{tikz}
\usepackage{dsfont}
\usetikzlibrary{calc,patterns,decorations.pathmorphing,decorations.markings,shapes,arrows}
\usepackage{parskip}
\usepackage{hyperref}
\usepackage{url}
\usepackage{parskip}
\usepackage{circuitikz}
\usepackage{nicefrac}
\usepackage[utf8]{inputenc}
\title{\LARGE \bf
Identification of optimal prediction error Thévenin models of Li-ion cells using the MOLI approach}

\author{P. Lopes dos Santos\thanks{P. Lopes dos Santos is with INESC TEC   \& FEUP,  Universidade do Porto, Portugal,   {\tt\small pjsantos@fe.up.pt}}%
,  T-P Azevedo Perdicoúlis\thanks{T-P Azevedo-Perdico\'{u}lis is with ISR---Coimbra \& Departamento de Matem\'atica, UTAD, 5001-801 Vila Real, Portugal        {\tt\small  tazevedo@utad.p}}  and Paulo A. Salgado\thanks{Paulo A. Salgado is with  Departamento de Engenharias, UTAD, 5001-801 Vila Real, Portugal        {\tt\small  psal@utad.pt}}%
\,\,\,\thanks{FCT - Funda\c c\~ao para a Ci\^{e}ncia e a Tecnologia under project:  (i) UIDB/50014/2020 for the first author.   (ii)  UIDB/00048/2020 for the second author. (iii) UIDB/04033/2020 for the third author.}
}

\begin{document}
\maketitle
\section{Introduction}\label{Introduction}
One of the challenges in designing battery management systems is to find a suitable model for its cells. Since it is not possible to guarantee that all cells are the same, it is convenient to estimate these models from data, using system identification algorithms.

This report starts by studying the dependence of OCV on SOC in Section~\ref{Opencircuitvoltage}.  In Section~\ref{Batteryequivalentseriesresistancemodel}, the  battery equivalent model when a resistor is added to the circuit is stated. 
As the discharge data is divided into segments where $C_{0},R_{0}$ are assumed constant, and therefore SOC is constant,   thence is described 
an  LTI identification algorithm to be used to estimate the cell model in each segment.    In Section~\ref{Randlescircuitdiffusionmodel}, 
the Randles  circuit diffusion model is described.  
   In particular, the Warburg impedance is discussed.   Also, after presenting the simplified Randles circuit, is stated an identification algorithm that estimates the parameters of this model.  
  In Section~\ref{ModifiedTheveninDm},   is enunciated an algorithm to identify a Thévenin model of 1st and 2nd order.   In Section~\ref{caseStudy}, the performance of the two models described in sections \ref{Randlescircuitdiffusionmodel} and \ref{ModifiedTheveninDm}, and its respective identification algorithms, is discussed and compared using an experimental set of data.
 
\section{Open circuit voltage}\label{Opencircuitvoltage}
A Li-Ion cell delivers a voltage at his terminals. If the cell is in open circuit, i.e, if there isn't any circuit connected to the cell, the voltage remains constant. Hence, it can be seen as a voltage source with a certain open-circuit voltage (OCV). It is known that the OCV of a fully charged cell is generally higher than the OCV of a discharged one. This can be included in the model by using a voltage source controlled by the state of charge (SOC) of the cell. The SOC is a dimensionless quantity that is 100\% (or 1) when the cell is fully charged and is 0\% (or 0) when it is fully discharged.  It is defined as
\begin{equation}
\label{SOC}
SOC(t)=\dfrac{Q_{max}-q_d(t)}{Q_{max}}\times 100\%=\left(1-\dfrac{q_d(t)}{Q_{max}}\right)\times 100\%,
\end{equation}
where $Q_{max}$ is the maximum charge the cell can store (cell capacity) and $q_d(t)$ is the charge removed from the cell. The $OCV$ is a function of $SOC$ and it is monotonous crescent as it
\begin{figure}[h!]
\begin{center}
\includegraphics[scale=1]{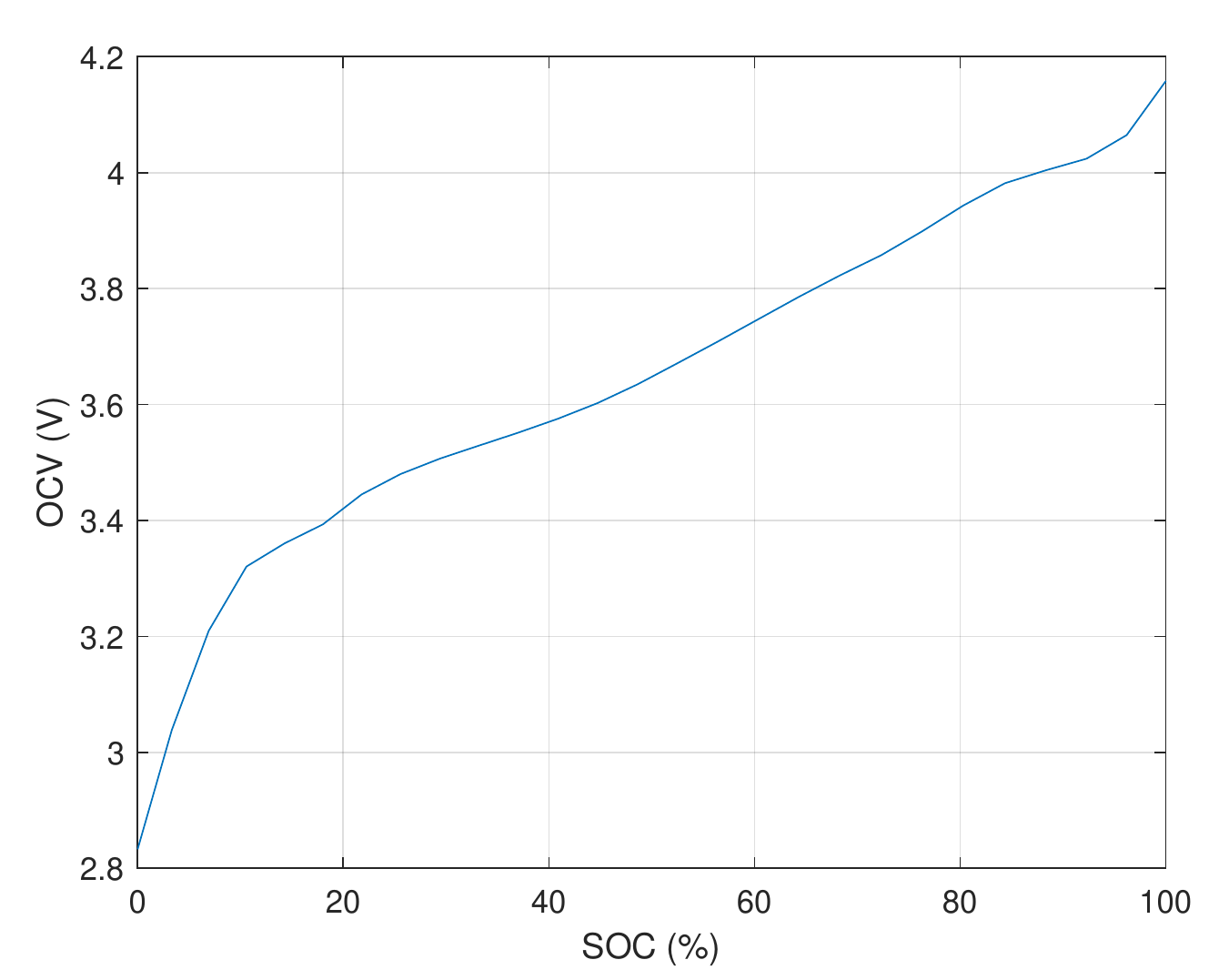}
\caption{OCV as a function of the SOC} 
\label{Fig1}
\end{center}
\end{figure}
can be seen in Figure \ref{Fig1}. If the cell is discharging, $q_d(t)$ is given by
\begin{equation}
\label{qd}
\dot{q}_d(t)=i_{bat}(t),
\end{equation}
where $i_{bat}(t)$ is the discharge current. If, at time instant $t$, an infinitesimal amount of charge $dq_d(t)$ is removed from the cell, the voltage at its terminals decreases by an amount of $dOCV(t)$, proporcional to $dq_d(t)$. Thus, we can write
\begin{equation}
\label{OCVandQ}
dOCV(t)=-\dfrac{1}{C_0(t)}dq_d(t).
\end{equation}  
Dividing this equation by $dt$ 
\begin{equation}\label{eq4}
\dfrac{dOCV(t)}{dt}=-\dfrac{1}{C_0(t)}\dfrac{dq_d(t)}{dt}=-\dfrac{1}{C_0(t)}i_{bat}(t),
\end{equation}
then the OCV can be seen as the voltage at the terminals of a time varying capacitor as depicted in Figure \ref{Fig2}.
\begin{figure}[h!]
\begin{center}
\input{OCV}
\end{center}
\caption{OCV as the voltage at the terminals of a time varying capacitor} 
\label{Fig2}
\end{figure}
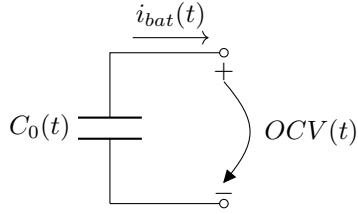
From equation \eqref{OCVandQ} 
\begin{equation}
\dfrac{dOCV}{dq_d}=-\dfrac{1}{C_0}.
\end{equation}
On the other hand, from \eqref{SOC},
\begin{equation}
dq_d=-\dfrac{Q_{max}}{100}dSOC
\end{equation}
whereby
\begin{equation}
\label{dOCVdSOC}
\dfrac{dOCV}{dSOC}=\dfrac{Q_{max}}{100C_0}.
\end{equation}
Therefore, from Figure \ref{Fig1} we see that $C_0$ is a function of the SOC, i.e, $C_0=C_0(SOC)$. 
\section{Battery equivalent series resistance model}\label{Batteryequivalentseriesresistancemodel}
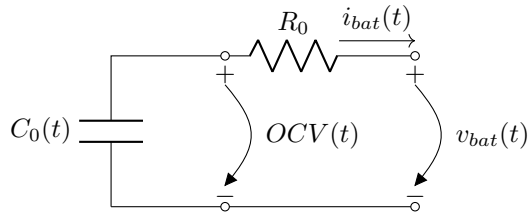
\begin{figure}[h!]
\begin{center}
\input{Rint}
\end{center}
\caption{Equivalent series resistance model} 
\label{Fig3}
\end{figure}

When a load is connected to the cell, its voltage drops. This can be modelled by a resistor in series with the capacitor as shown in Figure \ref{Fig3}.
This is the series resistance equivalent (SRE) model. With the SRE, recalling \eqref{eq4} and knowing that OCV depends on SOC, the cell model becomes
\begin{eqnarray}
\dot{OCV}(t)&=&-\dfrac{1}{C_0(SOC)}i_{bat}(t),\\
v_{bat}(t)&=&OCV(t)-R_0i_{bat}(t),
\end{eqnarray}
where $v_{bat}$ is the voltage at the cell terminals. This can be seen as a   continuous-time, quasi LPV state-space model:
\begin{eqnarray}
\dot{x}(t)&=&A_cx(t)+B_c(p_t)u(t),\\
y(t)&=&Cx(t)+Du(t),
\end{eqnarray}
with input $u(t)=i_{bat}(t)$, output $y(t)=v_{bat}(t)$, scheduling signal $p_t=SOC(t)$ and parameters $A_c=0$, $B_c(t)=-\dfrac{1}{C_0(t)}$, $C=1$ and $D=-R_{0}.$  In discrete-time, assuming that $i_{bat}(t)$ and $SOC(t)$ are constant between samples (ZOH digital to analog converters), the equivalent cell model is
\begin{eqnarray}
\label{OCVkp1}
OCV[k+1]&=&OCV[k]-\dfrac{T_s}{C_0(p_k)}i_{bat}[k]\\
\label{vbat1}
v_{bat}[k]&=&OCV[k]-R_0i_{bat}[k].
\end{eqnarray}
Also, from \eqref{qd} and \eqref{SOC},
\begin{eqnarray}
\label{qd1}
q_d[k+1]&=&q_d[k]+T_si_{bat}[k]\\
p_k=SOC[k]&=&\left(1-\dfrac{q_d[k]}{Q_{max}}\right)100. \label{etSOC}
\end{eqnarray}
When $k=0,$   \eqref{etSOC} becomes:
\begin{equation}
q_d[0]=\left(1-\dfrac{SOC[0]}{100}\right)Q_{max}.
\end{equation}
Note that $R_o$ is often a function of   SOC and   always a function of the temperature. In what follows, we assume  constant temperature   and    that both $R_0$ and $C_0$ are   piecewise constant functions of   SOC.   Consequently,   these parameters remain constant in the time intervals  where both $C_0$ and $R_0$ are constant (that is,   the intervals where SOC is constant).  For this reason, in each one of these intervals --- hereforth called segments ---  the cell model is   time invariant.  So,  to identify the piecewise LTI model,  the discharge data is divided into several segments, where every segment-i has $N_{i}$ data points. 
For each  segment,   an LTI identification algorithm is used with the OCV  initial value being the final    of the   previous one (except for the first one where the initial OCV  also  needs to be estimated). 
 
\subsection{Identification of the  series resistance model}\label{IdentificationSeriSREesistanceModel}
We   derive the LTI identification algorithm that will be used to identify the cell model in each segment-${i}$ containing $N_{i}$ data points. Using the shift forward operator $z$, i.e.,  $z x[k]=x[k+1]$, and considering $C_0(p_k)=C_{0_i}=constant$, equation \eqref{OCVkp1} may be written as
\begin{equation}
OCV[k]=-\dfrac{T_s}{C_{0_i}}\dfrac{i_{bat}[k]}{z-1}.
\end{equation}
But, from \eqref{qd1},
\begin{equation}
T_s\dfrac{i_{bat}[k]}{z-1}=q_d[k].
\end{equation}
Consequently,
\begin{equation}\label{ocv}
OCV[k]=OCV[0]-\dfrac{1}{C_{0_i}}q_d[k],
\end{equation}
and, substituting \eqref{ocv} into \eqref{vbat1},
\begin{equation}
\label{out}
y[k]:=v_{bat}[k]=OCV[0]-\dfrac{1}{C_{0_i}}q_d[k]-R_{0_i}i_{bat}[k].
\end{equation}
In the first segment, i.e, if $i=1,$ define:
\begin{equation}\label{regressorRmodel}
\varphi_1[k]:=
\begin{bmatrix}
1 & -q_d[k] &- i_{bat}[k]
\end{bmatrix}
\end{equation}
where $q_s[k]$ is given by \eqref{qd1} with $q_d[0]=0$, and
\begin{equation}\label{parameterRmodel}
\theta_1:=
\begin{bmatrix}
OCV[0] & \dfrac{1}{C}_{0_1} & R_{0_1}
\end{bmatrix}^{T},
\end{equation}
and using \eqref{regressorRmodel} and \eqref{parameterRmodel} rewrite \eqref{out} as:
\begin{equation}
\label{regression}
y[k]=\varphi_i[k]\theta_i,
\end{equation}
with $i=1$. Given $y[k]$ and $i_{bat}[k]$ for $k=0,\dots,N_1-1$,  the  LSE   of $\theta_1$ is calculated as:
\begin{equation}
\label{LSQ}
\hat{\theta}_i=\left(\Phi_i^T\Phi_i\right)^{-1}\Phi_i^TY_{i}
\end{equation}
with $i=1,$ and
\begin{eqnarray}
\Phi_i&:=&
\begin{bmatrix}
\varphi[0]^T & \varphi[1]^T & \cdots & \varphi[N-1]^T
\end{bmatrix}^T,\\
Y_i&:=&
\begin{bmatrix}
y[0] & y[1] & \cdots & y[N_i-1]
\end{bmatrix}^T.
\end{eqnarray}
In any other segment, i.e., $i=2,\ldots$, define $k'=k-\sum_{j=1}^{i-1}N_j+1$ and rewrite \eqref{out} as
\begin{equation}
\label{outi}
y[k']:=v_{bat}[k']-OCV[k'=0]=-\dfrac{1}{C_{0_i}}q_d[k']-R_{0_i}i_{bat}[k'],
\end{equation}
with $OCV[k'=0]=OCV\left[\sum_{j=1}^{i-1}N_j-1\right]$. This equation can also be rewritten as \eqref{regression} with $k$ replaced by $k'$ and
\begin{eqnarray}
\varphi_i[k']&:=&
\begin{bmatrix}
-q_d[k'] &- i_{bat}[k']
\end{bmatrix},\\
\theta_i&:=&
\begin{bmatrix}
\dfrac{1}{C}_{0_1} & R_{0_1}
\end{bmatrix}^{T},
\end{eqnarray}
 with the estimate of $\theta_i$ still being given by \eqref{LSQ}.

\section{Randles  circuit diffusion model}\label{Randlescircuitdiffusionmodel}
\label{RandlesSection}
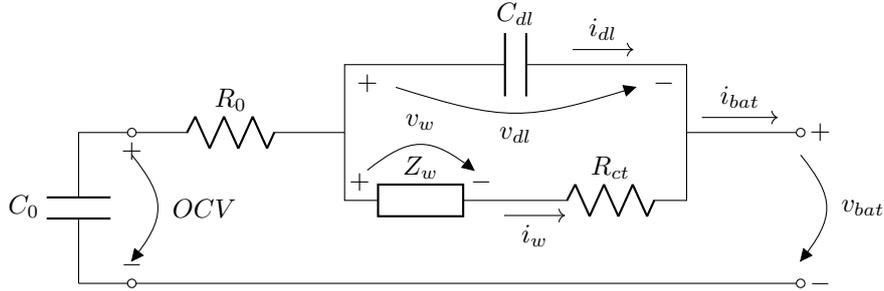
\begin{figure}[h!]
\begin{center}
\input{randlesCircuit}
\caption{Randles' circuit}
\label{Fig4}
\end{center}
\end{figure}

Cells are often modelled by the Randles circuit  depicted in Figure \ref{Fig4}. This circuit 
is inspired by electrochemical
principles   and it is recognised to be a trusty description of a cell dynamics  \cite{c19}.

Here, $R_0$ is the electrolyte resistance, $R_{ct}$ is the charge transfer resistance that models the voltage drop over the electrode–electrolyte interface due to a load, $C_{dl}$ is the double-layer capacitance
modelling the effect of charges building up in the electrolyte at the
electrode surface, and $Z_W$ is the so called Warburg impedance. The main difficulty is to model the Warburg impedance.

Next, the Warburg impedance is described and   discretised to be next approximated by a finite state-space realization.  To identify the parameters of the simplified Randles circuit, a MOLI like identification algorithm is formulated.     
\subsection{Warburg impedance}
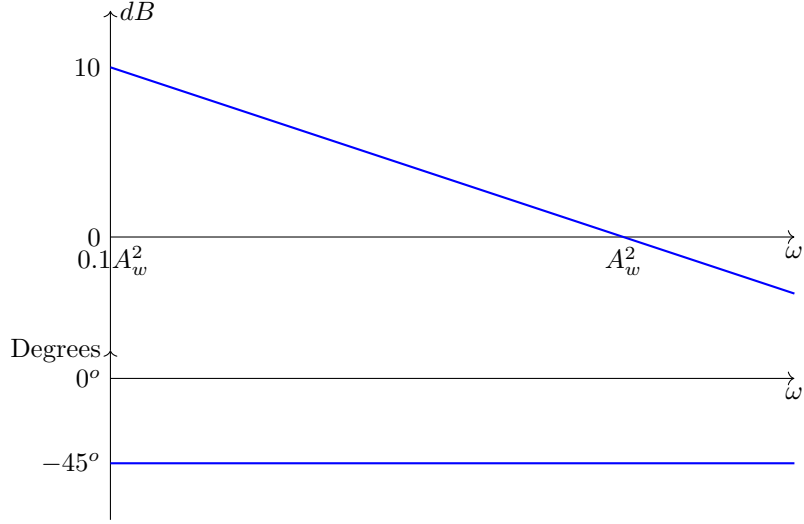
\begin{figure}[h!]
\begin{center}
\input{ZwBodeDiagrams} 
\caption{Bode diagrams of the Warburg impedance}
\label{Fig4b}
\end{center}
\end{figure}

The Warburg impedance models the diffusion of lithium ions in the electrodes. It is a frequency dependent impedance, given by
\begin{equation}
Z_w=\dfrac{A_w}{\sqrt{j\omega}},
\end{equation}
where $j=\sqrt{-1}$ is the imaginary unity,  $A_W$ is the Warburg coefficient, and $\omega$ is the frequency in radians per second. Figure \ref{Fig4b} shows the Bode diagrams of $Z_w,$ where it can be seen that the amplitude diagram is a straight line with a slope of $-10$  dB per decade, and the Phase   constant and equal to $-45^o$.
\subsubsection{Fractional integrator}
The Warburg impedance is a semi-integrator of the current. The semi-integrator is a special case of the fracional integrator of order $\alpha$ with transfer function $\dfrac{1}{s^{\alpha}}$. It is well known that the inverse Laplace transform of $\dfrac{1}{s^n}$ is zero for $t<0$ and $\dfrac{t^{n-1}}{(n-1)!}$ for $t\geq 0$, i.e.,
\begin{equation}
\mathcal{L}^{-1}\left\{\dfrac{1}{s^n}\right\}=
\left\{
\begin{array}{ll}
0,& t<0\\
\dfrac{t^{n-1}}{(n-1)!}, & t\geq 0
\end{array}
\right.
0\dfrac{t^{n-1}}{\Gamma(n)}\boldsymbol{1}(t),
\end{equation}
where $\boldsymbol{1}(t)$ is the unit step. As for any positive integer $n$ the Gamma function is given by,
\begin{equation}
\Gamma(n)=\displaystyle	\int_{0}^{\infty}x^{n-1}e^{-x}dx=(n-1)!,
\end{equation} 
then
\begin{equation}
\mathcal{L}^{-1}\left\{\dfrac{1}{s^n}\right\}=
\left\{
\begin{array}{ll}
0,& t<0\\
\dfrac{t^{n-1}}{\Gamma(n)},& t\geq 0
\end{array}
\right.
=\dfrac{t^{n-1}}{\Gamma(n)}\boldsymbol{1}(t).
\end{equation}
Generalizing this result for $\mathcal{L}^{-1}\left\{\dfrac{1}{s^{\alpha}}\right\}$, with $\alpha \in\mathds{R}^+$, yields,
\begin{equation}
\mathcal{L}^{-1}\left\{\dfrac{1}{s^\alpha}\right\}=
\left\{
\begin{array}{ll}
0,& t<0\\
\dfrac{t^{\alpha-1}}{\Gamma(\alpha)},& t\geq 0
\end{array}
\right.
\dfrac{t^{n-1}}{\Gamma(n)}\boldsymbol{1}(t).
\end{equation}
This relation can be confirmed by the calculation of $\mathcal{L}\left\{t^{\alpha-1}\boldsymbol{1}(t)\right\}$:
\begin{equation}
\mathcal{L}\left\{t^{\alpha-1}\boldsymbol{1}(t)\right\}=\displaystyle\int_{0}^{\infty}t^{\alpha-1}e^{-st}dt=\int_{0}^{\infty}\dfrac{u^{\alpha-1}}{s^{\alpha-1}}e^{-st}\dfrac{du}{s}=\dfrac{1}{s^{\alpha}}\int_{0}^{\infty}u^{\alpha-1}e^{-u}du=\dfrac{\Gamma(\alpha)}{s^{\alpha}}.
\end{equation}
Therefore,
\begin{equation}
\mathcal{L}\left\{\dfrac{t^{\alpha-1}\boldsymbol{1}(t)}{\Gamma(\alpha)}\right\}=\dfrac{1}{s^{\alpha}}\Leftrightarrow\mathcal{L}^{-1}\left\{\dfrac{1}{s^{\alpha}}\right\}=\dfrac{t^{\alpha-1}}{\Gamma(\alpha)}\boldsymbol{1}(t).
\end{equation}
\subsection{Impulse response of the sampled Warburg impedance}
The system is sampled with a Zero Order Hold (ZOH)  to obtain the discrete time system, followed by its rational approximation.
\subsubsection{Zero Order Hold sampling}
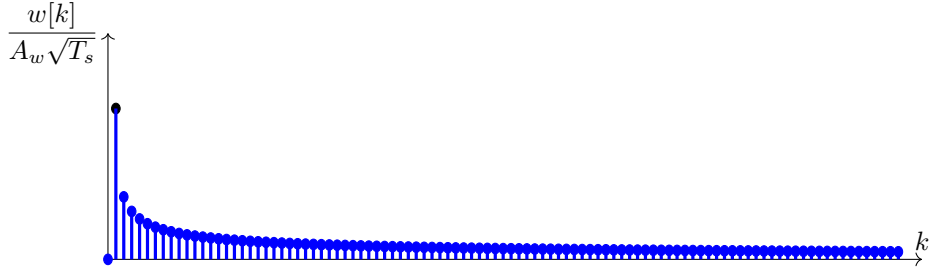
\begin{figure}[h!]
\begin{center}
\input{WarburgImpulse}
\caption{Normalized Discrete-time impulse response of the Warburg Impedance}
\label{Fig5}
\end{center}
\end{figure}

The output of a fractional integrator is
\begin{equation}
y(t)=\displaystyle\int_{\tau=0}^{\infty}\dfrac{\tau^{\alpha-1}}{\Gamma(\alpha)}u(t-\tau)d\tau.
\end{equation}
If $u(t)$ is the output of a ZOH system (p.e, a DA converter), then it is constant between to consecutive sampling instants, i.e.,
\begin{equation}
u(t)=u(kT_s)=u[k], kT_s\leq t < (k+1)T_s,
\end{equation}
where $T_s$ is the sampling period. For this input, the output at the sampling instant $t=kT_s$ is
\begin{eqnarray}
\nonumber
y(kT_s)&=&y[k]=\dfrac{1}{\Gamma(\alpha)}\displaystyle\sum_{\ell=1}^{\infty}\displaystyle\int_{(\ell-1) T_s}^{\ell Ts}\tau^{\alpha-1}u[k-\ell]d\tau \\
\nonumber
&=&\dfrac{1}{\alpha\Gamma(\alpha)}\sum_{\ell=1}^{\infty}\left[\tau^{\alpha}\right]_{\tau=(\ell-1) T_s}^{\ell T_s}u[k-\ell]=\displaystyle\sum_{\ell=1}^{\infty}\dfrac{\left(\ell T_s\right)^{\alpha}-\left((\ell-1) T_s\right)^{\alpha}}{\alpha\Gamma(\alpha)}u[k-\ell] \\
\nonumber
&=&
\dfrac{T_s^{\alpha}}{\alpha\Gamma(\alpha)}\sum_{\ell=1}^{\infty}\left(\ell^{\alpha}-(\ell-1)^{\alpha}\right)u[k-\ell]=\dfrac{T_s^{\alpha}}{\alpha\Gamma(\alpha)}\left(h[k]*u[k]\right),
\end{eqnarray}
where $*$ stands for convolution and
\begin{equation}
h[k]=k^{\alpha}-(k-1)^{\alpha},\quad k=1,2,\dots,\infty.
\end{equation}
Hence, the impulse response of the fractional integrator is
\begin{equation}
h[k]=
\left\{
\begin{array}{ll}
0,& k<1 \\
\dfrac{T_s^{\alpha}}{\alpha \Gamma(\alpha)}\left(k^{\alpha}-(k-1)^{\alpha}\right),& k\geq 1
\end{array}
\right.
=\dfrac{T_s^{\alpha}}{\alpha \Gamma(\alpha)}\left((k+1)^{\alpha}-k^{\alpha}\right)\boldsymbol{1}[k-1],
\end{equation}
where $\boldsymbol{1}[k]$ is the discrete-time unit step. As the Warburg impedance is a fractional integrator with $\alpha=0.5$, its impulse response is
\begin{eqnarray}
\nonumber
w[k]&=&
\left\{
\begin{array}{ll}
0,& k<1 \\
\dfrac{2A_{\omega}\sqrt{T_s}}{\Gamma(0.5)}\left(\sqrt{k}-\sqrt{k-1}\right),&k\geq 1
\end{array}
\right.\\[2pt]
&=&
\dfrac{2A_w{\omega}\sqrt{T_s}}{\Gamma(0.5)}\left(\sqrt{k}-\sqrt{k-1}\right)\boldsymbol{1}
[k-1]\\ 
&=& 1.1284A_{\omega}\sqrt{T_s}\left(\sqrt{k}-\sqrt{k-1}\right).
\end{eqnarray}
Figure \ref{Fig5} shows the normalised impulse response of the Warburg impedance, $\dfrac{w[k]}{A_{\omega}\sqrt{T_s}}.$ 

\subsubsection{Rational approximation of the Warburg impedance}
\label{warburgImpedance}
\begin{figure}[h!]
\begin{center}
\input{compareWarburg}
\caption{Discrete-time impulse responses of the Warburg Impedance (in blue) and its rational approximation (in red)}
\label{Fig6}
\end{center}
\end{figure}
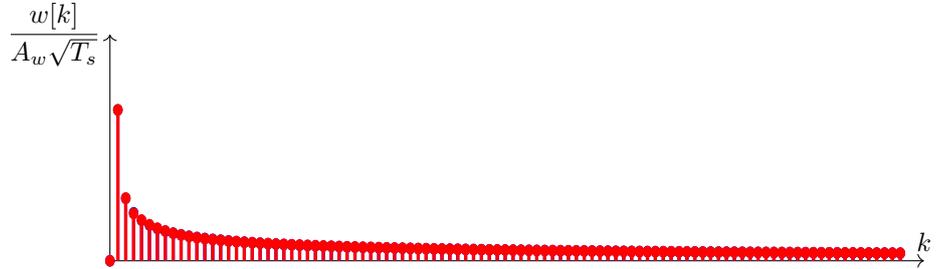

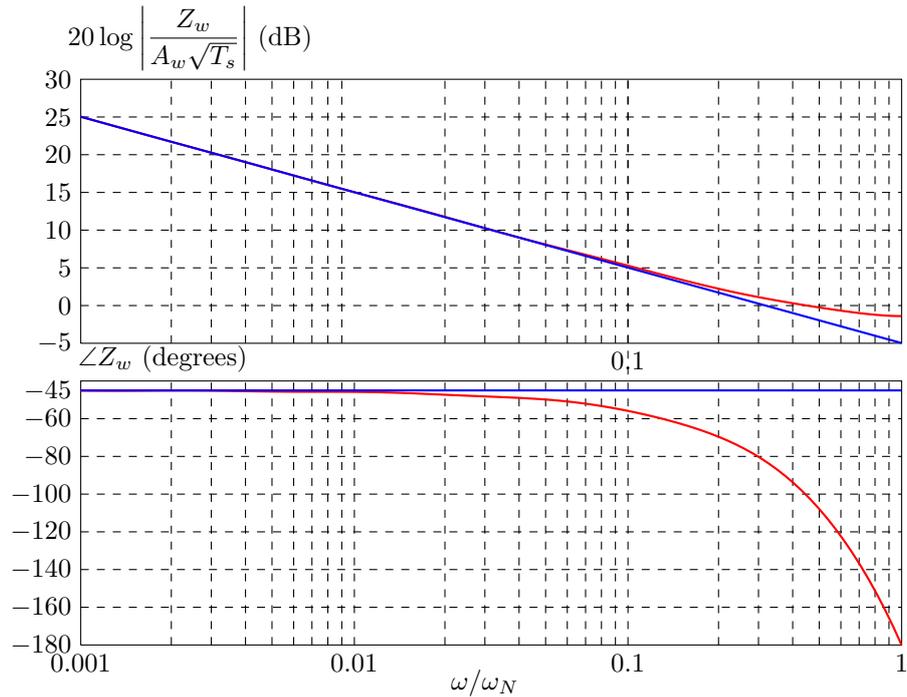
\begin{figure}[h!]
\begin{center}
\input{bodePlots}
\caption{Bode plots the Warburg Impedance (in blue) and its rational approximation (in red)}
\label{Fig7}
\end{center}
\end{figure}

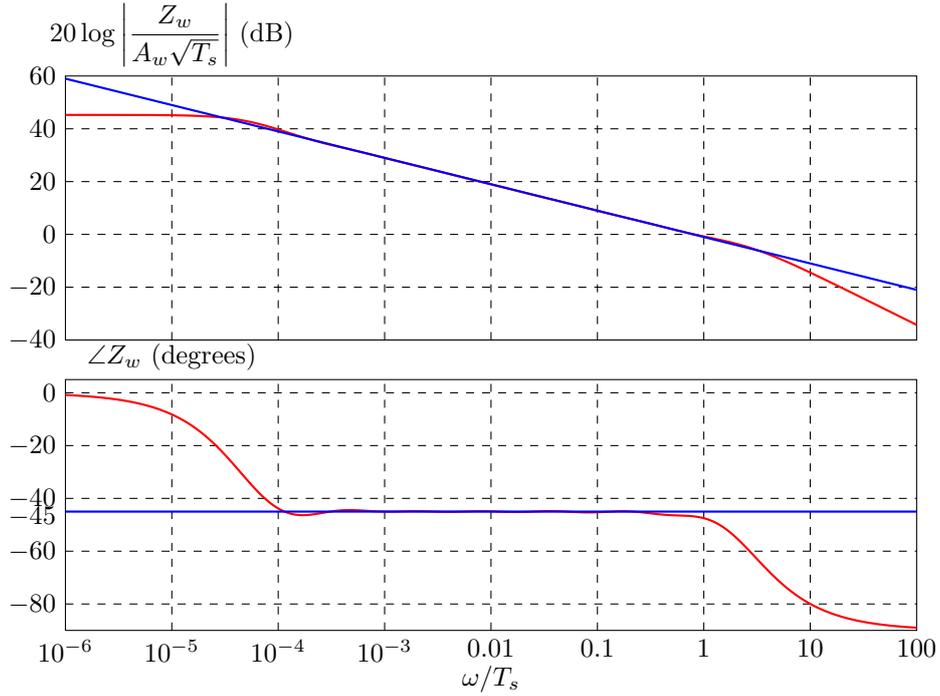
\begin{figure}[h!]
\begin{center}
\input{bodePlotsC}
\caption{Bode plots the Warburg Impedance (in blue) and its continuous-time rational approximation (in red)}
\label{Fig7C}
\end{center}
\end{figure}

The discrete-time normalized Warburg impedance $\dfrac{Z_w}{A_w\sqrt{T_s}}$ was approximated by the following state-space realization of a rational transfer function using the Ho-Kalman algorithm.
\begin{eqnarray}
\label{Wx}
x_w[k+1]&=&A_zx_w[k]+B_zi_w[k]\\
\label{Wy}
y_w[k]&=&C_zx_w[k]
\end{eqnarray}
with
\begin{eqnarray}
\nonumber
A_z&\hspace{-5mm}=&
\hspace{-5mm}
\footnotesize{
\begin{bmatrix}
 0.99964    & -0.0014121  & -0.0025413 & -0.0028264 & -0.0012488 &  
 0.00041095 &  0.00012636 \\
-0.0014120  &  0.98934     & -0.028798  & -0.036660  & -0.018323  &   
 0.0065444  &  0.0023842  \\
-0.0025412  & -0.028798    &  0.88467  & -0.18947   & -0.11661   &     
 0.051034   &  0.021731   \\
-0.0028263  & -0.036660    & -0.18947   &  0.61030   & -0.30015   & 
 0.16198    &  0.080232   \\
-0.0012488  & -0.018323    & -0.11661   & -0.30015   &  0.70033   &   
 0.21261    &  0.12742    \\
 0.00041094 &  0.0065444   &  0.051034  &  0.16198   &  0.21261   &
 0.78626    & -0.17239    \\
 0.00012635 &  0.0023842   &  0.021731  &  0.080232  &  0.12742   & -0.17239    &  0.80393
\end{bmatrix}}\\
\label{Az}
\\
\label{Bz}
B_z&\hspace{-5mm}=&
\hspace{-5mm}
\footnotesize{
\begin{bmatrix}
 0.194140 & 0.376185 & 0.631849 & 0.673511 & 0.294717 & -0.092444 & -0.029057
\end{bmatrix}^T}\\
\label{Cz}
C_z&\hspace{-5mm}=&
\hspace{-5mm}
\footnotesize{
\begin{bmatrix}
 0.194143 & 0.376185 & 0.631849 & 0.673511 & 0.294717 & -0.092444 & -0.029057
\end{bmatrix}}
\end{eqnarray}
This approximation has a relative error of
\begin{equation}
E_{10000}=0.45\%,
\end{equation}
where
\begin{equation}
E_T=e_{T_{rms}}/w_{T_{rms}}*100\%,
\end{equation}
with $e_{T_{rms}}$ and $w_{T_{rms}}$  being the rms values of 
\begin{equation}
e[0:T]=w[0:T]-\hat{w}[0:T]
\end{equation}
and $w_{0:T}$. Here, $w_{0:T}$ and $ \hat{w}_{0:T}$ are vectors with the samples for $k=0$  to $k=T$ of the impulse responses $w[k]$ and $\hat{w}[k]$ of the Warburg impedance and its rational approximation, respectively. In Figure \ref{Fig6}, it cannot be seen any difference between the impulse  responses of the Warburg impedance and its ractional approximation because they completely overlap.
Figure \ref{Fig7} compares the Bode plots where it can be seen that the match is almost perfect up to a frequency of $0.02\omega_N$ where $\omega_n$ is the Nyquist frequency.
The continuous time rational approximations of Warburg impedance can be derived from this discrete-time approximation, beeing equal to
\begin{eqnarray}
\bar{A}_{z}&=&\dfrac{1}{T_s}\ln\left(A_z\right)=\\\nonumber
&&
\hspace{-15mm}
\footnotesize{\begin{bmatrix}
-0.3835 &-1.7084  &  -4.0692 &  -6.1460 &  -4.1516 &   2.3824 &  1.3965\\
-1.7082 &-14.4359 & -48.5062 & -79.8499 & -56.4056 &  32.6265 & 19.2953\\
-4.0691 &-48.5062 &-219.8206 &-421.5413 &-324.2213 & 195.5203 & 116.6362\\
-6.1458 &-79.8499 &-421.5413 &-911.8301 &-774.4758 & 498.2598 & 304.6979\\
-4.1515 &-56.4056 &-324.2213 &-774.4758 &-739.1521 & 532.3045 &  345.9606\\
 2.3823 & 32.6265 & 195.5203 & 498.2598 & 532.3045 &-454.9094 &  -343.3191\\
 1.3964 & 19.2953 & 116.6362 & 304.6979 & 345.9606 &-343.3191 & -321.9213
\end{bmatrix}
\dfrac{10^{-3}}{T_s}}\\
\bar{B}_z&=&-\left(I-A_z\right)^{-1}\bar{A}_zB_z=\\
\nonumber
&=&
\dfrac{1}{T_s}
\begin{bmatrix}
 0.1881 & 0.4215 & 0.9410 & 1.3845 &  0.9325 & -0.5311 6 -0.3131
\end{bmatrix}\\
\bar{C}_z&=&C_w=\\
\nonumber
&=&
\footnotesize{
\begin{bmatrix}
 0.194143 & 0.376185 & 0.631849 & 0.673511 & 0.294717 & -0.092444 & -0.029057
\end{bmatrix}}.
\end{eqnarray} 
Figure \ref{Fig7C} compares the Bode plots of $Z_w$ and its continuous-time rational approximation. The match is almost perfect for two decades($10^{-4}/T_s<\omega<1/T_s$).
\subsection{Identification of the Randles' circuit parameters}
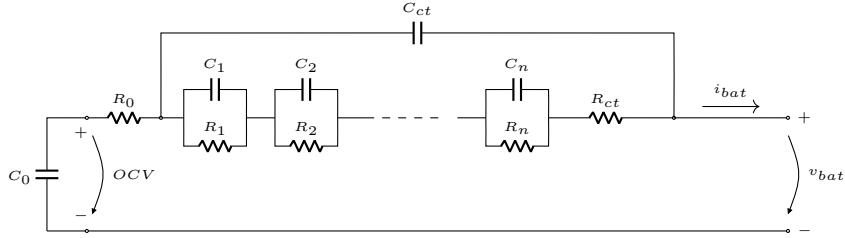
\begin{figure}[h!]
\begin{center}
\input{seriesRCParallell}
\caption{High order approximation of the Warburg impedance}
\label{Fig8}
\end{center}
\end{figure}

 Usually, this impedance is approximated by several series of RC parallel circuits, leading to the battery  equivalent circuit depicted in Figure \ref{Fig8}. 
It was seen in the previous section that the Warburg impedance can be approximated by a $7^{th}$ order LTI system which means that the circuit in Figure \ref{Fig8} must have at least 7 RC parallels to achieve this approximation, i.e, $n$ must be set to 7 and a naive approach leads to a dynamic system with 17 parameters to be identified. But the approximation of the Warburg impedance, derived in the previous section, has only one unknown parameter.  Therefore, using this approximation as a priori knowledge reduces the number of unknown parameters to five:
$C_0$,
$C_{ct}$,
$R_o$,
$R_{ct}$ and 
$A_w$.
This number can be further reduced to three, since often the double layer capacitance, $C_{ct}$, is negligible, and when this happens, the charge transfer resistance,$R_{ct}$, and the electrolyte resistance, $R_o$, are joined into a single resistance equal to $R_{ct}+R_o$.\\
In this section an algorithm  is formulated to estimate  the parameters of the simplified Randles' circuits (without the double layer capacitance). 

In a similar manner to Subsection~\ref{IdentificationSeriSREesistanceModel}, 
this  algorithm  do not assume variability in the  parameters. Instead,  it is  used in several segments of the discharge processes to estimate a piecewise LTI model.     
\subsubsection{Simplified Randles' circuit}\label{simplifiedRC}
\label{3.1}
\begin{figure}[h!]
\begin{center}
\input{simplifiedRandlesCircuit}
\caption{Simplified Randles' circuit}
\label{Fig9}
\end{center}
\end{figure}
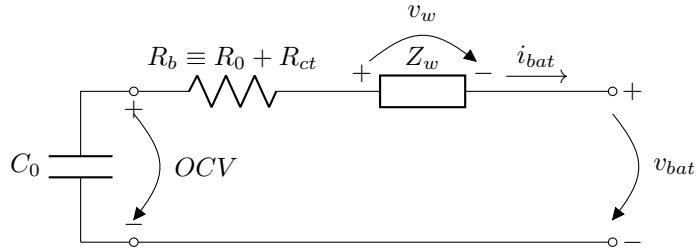

The simplified Randles' circuit  depicted in Figure \ref{Fig9} is
approximated by the following equations
\begin{eqnarray}
\dot{OCV}(t)&=&-\dfrac{1}{C_0}i_{bat}(t),\\
\dot{x}_w(t)&=&\bar{A}_zx_w(t)+\bar{B}_zi_{bat}(t),\\
v_{bat}(t)&=&OCV(t)-A_w\sqrt{T}_s\bar{C}_{z}x_w(t)-R_b i_{bat}(t),
\end{eqnarray}
where $A_w$ is the Warburg coefficient and $(\bar{A}_z,\bar{B}_z,A_w\bar{C}_z)$ is a
realisation of a continuous-time approximation of the Warburg impedance. The correspondent ZOH discrete-time model is
\begin{eqnarray}
\label{voc}
OCV[k+1]&=&OCV[k]-\frac{T_s}{C_0}i_{bat}[k],\\
\label{xw}
x_{w}[k+1]&=&A_zx_w[k]+B_zi_{bat}[k],\\
\label{vbat}
v_{bat}[k]&=&OCV[k]-A_w\sqrt{T_s}C_zx_w[k]-R_bi_{bat}[k],
\end{eqnarray}
where $T_s$ is the sampling period and $(A_z,B_z,A_w\sqrt{T_s}C_w,0)$ is the discrete-time realization of the Warburg matrices with $A_z$, $B_z$ and $C_z$ given in equations \eqref{Az}-\eqref{Cz}. Equation \eqref{voc} yields \eqref{OCVkp1}, that is:
\begin{equation}
\label{voc1}
OCV[k]=OCV[0]-\dfrac{T_s}{C_0}\sum_{\tau=0}^{k}i_{bat}[\tau]=OCV[0]-\dfrac{1}{C_0}q_{d}[k].
\end{equation}

On the other hand, using the forward time shift operator in equation \eqref{xw} 
\begin{equation}
\label{xw1}
x_w[k]=\left(zI-A_z\right)^{-1}B_zi_{bat}[k]=A_z^kx_w[0]+x_{w0}[k],
\end{equation}
where $x_{w0}[k]$ is the state  the output of the system $(A_z,B_z,I)$ driven by $i_{bat}[k]$ with zero initial state. Using \eqref{voc1} and \eqref{xw1} in \eqref{vbat}, the following regressor is obtained for $i=1$
\begin{eqnarray*}
\nonumber
y[k]:=v_{bat}[k]&=&OCV[0]-C_zA_z^kx_{w}[0]\sqrt{T_s}A_w-q_{d}[k]\dfrac{1}{C_0}\\
&&-C_zx_{w0}[k]\sqrt{T_s}A_w-R_bi_{bat}[k] \\
&=&
\begin{bmatrix}
1 & -C_zA_z^k & -q_{d}[k] & -C_zx_{w0}[k] & i_{bat}
\end{bmatrix}
\begin{bmatrix}
OCV[0] \\ \sqrt{T_s}A_wx_{w}[0] \\ \dfrac{1}{C_0}\\ \sqrt{T_s}A_w \\ R_b
\end{bmatrix}\\ 
&=&\varphi_{1}[k]\theta_{1}
\end{eqnarray*}
where
\begin{eqnarray}
\varphi_{1}[k]&:=&
\begin{bmatrix}
1 & -C_zA_z^k & -q_{d}[k] & -C_zx_{w0}[k] & i_{bat}
\end{bmatrix}\\
\theta_{1}&:=&
\begin{bmatrix}
OCV[0] & \sqrt{T_s}A_wx_{w}[0] & \dfrac{1}{C_0} & \sqrt{T_s}A_w & R_b
\end{bmatrix}^T.
\end{eqnarray}

In a similar manner as in Subsection~\ref{IdentificationSeriSREesistanceModel}, for $i=2,\ldots $ we have:
\begin{eqnarray}
\nonumber
y[k]& := &v_{bat}[k]-OCV[0] \\
&=&  -q_{d}[k]\dfrac{1}{C_0} - \left( C_zA_z^kx_{w}[0]  +C_zx_{w0}[k]\right)\sqrt{T_s} A_w -R_bi_{bat}[k]=\\
&=&\begin{bmatrix}
 -q_{d}[k] & -C_zA_z^k  -C_zx_{w0}[k] & i_{bat}
\end{bmatrix}
\begin{bmatrix}
\dfrac{1}{C_0}\\  \sqrt{T_s}A_w \\ R_b
\end{bmatrix}
\\&=& \varphi_{i}[k]\theta_{i},
\end{eqnarray}
where
\begin{eqnarray}
\varphi_{i}[k]&:=&
\begin{bmatrix}
 -q_{d}[k] & -C_zA_z^k  -C_zx_{w0}[k] & i_{bat}
\end{bmatrix},\\
\theta_{i}&:=&
\begin{bmatrix}
\dfrac{1}{C_0} & \sqrt{T_s}A_w & R_b
\end{bmatrix}^T.
\end{eqnarray}
Defining
\begin{eqnarray}
Y_i&:=&
\begin{bmatrix}
y[0] & y[1] & \cdots &y[N_{i}-1]
\end{bmatrix}^T, \quad i=1,2, \ldots
\\
\Phi_i&:=&
\begin{bmatrix}
\varphi_{i}[0]^T & \varphi_{i}[1]^T & \cdots & \varphi_{i}[N_{i}-1]^T 
\end{bmatrix}^T,  \quad i=1,2, \ldots
\end{eqnarray}
then
\begin{equation}
Y_i=\Phi_i\theta_{i},
\end{equation}
and if $\Phi_i$ is a full rank matrix, $\theta$ can be estimated by the  LSE:
\begin{equation}
\label{SimpleRandleLSQ}
\hat{\theta}_i=\Phi_i^{\dag}Y_i=\left(\Phi_i^T\Phi_i\right)^{-1}\Phi_i^TY_i.
\end{equation}

\section{The Modified Thévenin Difusion model}\label{ModifiedTheveninDm}
\begin{figure}[h!]
\begin{center}
\input{seriesRCParallell0}
\caption{High order equivalent circuit}
\label{Fig4c}
\end{center}
\end{figure}
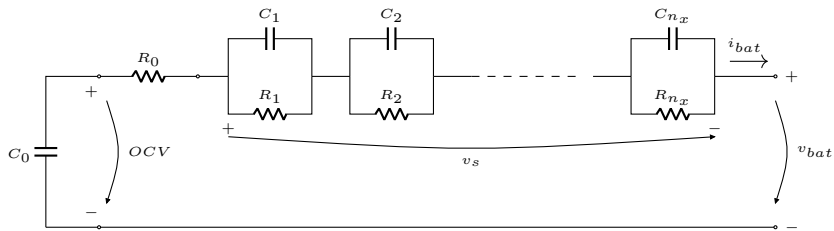

In this section  the battery is modelled by the circuit of Figure \ref{Fig4c}, 
which can be described by the state-space model
\begin{eqnarray}
\label{MS1}
\dot{OCV}(t)&=&-\dfrac{1}{C_0}i_{bat}(t),\\
\label{MS2}
\dot{x}(t)&=&A_cx(t)+B_c i_{bat}(t),\\
\label{MS3}
v_{bat}(t)&=&OCV(t)-Cx(t)-R_o i_{bat}(t),
\end{eqnarray}
where
\begin{eqnarray}
\dot{x}(t)&=&A_cx(t)+B_ci_{bat}(t),\label{eq1SS}\\
v_s(t)&=&Cx(t),\label{eq2SS}
\end{eqnarray}
describes the series of the RC circuits with
 input $i_{bat}(t)$, output $v_s(t)$. 
  If we specify the initial value of $OCV(0)$ in equation \eqref{MS3},
\begin{equation}
\label{MS3Iint}
v_{bat}(t)=OCV(0)-\dfrac{1}{C_0}q_{out}(t)-Cx-R_0 i_{bat}(t)
\end{equation}
where
\begin{equation}
\dot{q}_{out}(t)=i_{bat}(t),
\end{equation}
and define  next  the extended input
\begin{equation}
u_{e}(t)=
\begin{bmatrix}
\boldsymbol{1}(t) \\ i_{bat}(t) \\ q_{out}(t)
\end{bmatrix},
\end{equation}
with $\boldsymbol{1}(t)$ being the continuous-time unit step, we can rewrite \eqref{MS2} and \eqref{MS3Iint} as
\begin{eqnarray}
\dot{x}(t)&=&A_cx(t)+B_{ce}u_e(t)\\
v_{bat}(t)&=&Cx(t)+D_e u_e(t) \label{vout}
\end{eqnarray}
where some matrices need to be redefined:
\begin{equation}
B_{ce}:=
\begin{bmatrix}
0_n & 0_n & B_c
\end{bmatrix},
\qquad
D_{e}:=
\begin{bmatrix}
OCV(0) & -\dfrac{1}{C_0} & -R_0,
\end{bmatrix}.
\end{equation}
The next step is to   discretise the system. Then
\begin{eqnarray}
\label{Md1}
x[k+1]&=&Ax[k]+B_{e}u_e[k],\\
\label{Md2}
v_{bat}[k]&=&Cx[k]+D_e u_e[k],
\end{eqnarray}
where
\begin{eqnarray}
B_e&=&
\begin{bmatrix}
0_n & 0_n & B 
\end{bmatrix},\\
u_e[k]&=&
\begin{bmatrix}
\boldsymbol{1}[k] & q_{out}[k] &i_{bat}[k]
\end{bmatrix}^T 
\end{eqnarray}
and 
\begin{equation}
q_{out}[k+1]=q_{out}[k]+T_si_{bat}[k].
\end{equation}
Here $T_s$ is the sampling period.
\subsection{Identification algorithm}
Since the data has been partitioned into different segments, we identify a piecewise LTI model in every segment  in a similar manner as we have done for the other models. 
\subsubsection{Segment 1 -Unknown initial OCV and initial state} 
\label{Segment1}
We use a MOLI approach to formulate the identification algorithm, taking into account that both  the initial OCV and the initial state are unknown for the first segment.   
 To do this,  first partition $A$ as
\begin{equation}
A=A_0+LC.
\end{equation}
Then
\begin{equation}\label{eq90}
x[k+1]=A_0x[k]+LCx[k]+B_eu_e[k].
\end{equation}
From  \eqref{Md2},
\begin{equation}\label{eq91}
Cx[k]=v_{bat}[k]-D_eu_e[k].
\end{equation}
Substituting \eqref{eq91} into \eqref{eq90} becomes: 
\begin{equation}
x[k+1]=A_0x[k]+Lv_{bat}[k]+\left(B_e-LD_e\right)u_e[k],
\end{equation}
and  consequently 
\begin{equation}
x[k]=A_0^kx[0]+\left(qI-A_0\right)^{-1}Lv_{bat}[k]+\left(qI-A_0\right)^{-1}\left(B_e-LD_e\right)u_e[k].
\end{equation}
Expanding $B_e$, $D_e$ and $u_e[k]$,
\begin{eqnarray}
x[k]&=&A_0^kx[0]+\left(qI-A_0\right)^{-1}Lv_{bat}[k]+\nonumber\\
\nonumber
& &+\left(qI-A_0\right)^{-1}\left(
\begin{bmatrix}
0_n & 0_n & B
\end{bmatrix}
-L
\begin{bmatrix}
OCV[0] & -\dfrac{1}{C_0} & -R_0
\end{bmatrix}
\right)
\begin{bmatrix}
\boldsymbol{1}[k] \\ q_{out}[k] \\ i_{bat}[k]
\end{bmatrix} \\
\nonumber
&=& A_0^kx[0]+\left(qI-A_0\right)^{-1}Lv_{bat}[k]-\left(qI-A_0\right)^{-1}L\;OCV[0]\boldsymbol{1}[k]+\\
& &+\left(qI-A_0\right)^{-1}\dfrac{L}{C_0}q_{out}[k]+
\left(qI-A_0\right)^{-1}B_ai_{bat}[k]  \label{eq93}
\end{eqnarray}
with
\begin{equation}
B_a=B+ L R_0.
\end{equation}
Then substituting \eqref{eq93} into \eqref{vout}, we have:
\begin{eqnarray}
\nonumber
v_{bat}[k]&=&CA_0^kx[0]+C\left(qI-A_0\right)^{-1}Lv_{bat}[k]-C\left(qI-A_0\right)^{-1}L\; OCV[0]\boldsymbol{1}[k]+\\
\nonumber
& &+C\left(qI-A_0\right)^{-1}\dfrac{L}{C_0}q_{out}[k]+
C\left(qI-A_0\right)^{-1}B_ai_{bat}[k]+\\
\nonumber
& & +OCV[0]\boldsymbol{1}[k]-\dfrac{1}{C_0}q_{out}[k]-R_0i_{bat}[k]\\
\nonumber
&=&CA_0^kx[0]+C\left(qI-A_0\right)^{-1}i_{bat}[k]B_a-i_{bat}[k]R_0+\\
\nonumber
& &+C\left(qI-A_0\right)^{-1}v_{bat}[k]L+\boldsymbol{1}[k]OCV[0]-q_{out}[k]\dfrac{1}{C_0}-\\
& &-C\left(qI-A_0\right)^{-1}\boldsymbol{1}[k]L\; OCV[0]+C\left(qI-A_0\right)^{-1}q_{out}[k]\dfrac{L}{C_0}
\label{outMs1}
\end{eqnarray}
Defining
\begin{eqnarray*}
i_{Fbat}[k]&:=&\left(qI-A_0^T\right)^{-1}C^Ti_{bat}[k],\\
v_{Fbat}[k]&:=&\left(qI-A_0^T\right)^{-1}C^Tv_{bat[k]},\\
\boldsymbol{1}_{F}[k]&:=&\left(qI-A_0^T\right)^{-1}C^T\boldsymbol{1}[k],\\
q_{outF}[k]&:=&\left(qI-A_0^T\right)^{-1}C^Tq_{out}[k],
\end{eqnarray*}
where $\boldsymbol{1}[k]$ is the unit step, we can rewrite \eqref{outMs1} as
\begin{eqnarray}
\nonumber
v_{bat}[k]&=&CA_0^kx[0]+i_{Fbat}[k]^TB_a-i_{bat}[k]R_0+v_{Fbat}^T[k]L+\boldsymbol{1}[k]OCV[0]-\\
& &-q_{out}[k]\dfrac{1}{C_0}-\boldsymbol{1}_{F}^T[k]L\; OCV[0]+q_{outF}^T[k]\dfrac{L}{C_0}.\label{eq97}
\end{eqnarray}
Defining 
\begin{eqnarray*}
\theta_A&:=&x[0]\in\mathds{R}^{n_x}, \\
\theta_B&:=& B_a\in\mathds{R}^{n_x}, \\
\theta_C&:=&R_0\in\mathds{R}, \\
\theta_D&:=&L\in\mathds{R}^{n_x}, \\
\theta_E&:=&OCV[0]\in\mathds{R},\\
\theta_F&:=&\dfrac{1}{C_0}\in\mathds{R},\\
\theta_G&:=&L\; OCV[0]\in\mathds{R}^{n_x} ,\\
\theta_H&:=&\dfrac{L}{C_0}\in\mathds{R}^{n_x}. \\
\end{eqnarray*}
Observe that 
\begin{eqnarray}
\label{R1}
\theta_G&=&\theta_D\theta_E,\\
\label{R2}
\theta_H&=&\theta_D\theta_F.
\end{eqnarray}
Thence
\begin{eqnarray}
y[k]:=v_{bat}[k]&=&CA_0^K\theta_A+i_{Fbat}^T[k]\theta_B-i_{bat}[k]\theta_C+v_{Fbat}^T[k]\theta_D+\boldsymbol{1}[k]\theta_E  \nonumber \\
\nonumber
& &-q_{out}[k]\theta_F-\boldsymbol{1}_F^T[k]\theta_G+q_{outF}^T[k]\theta_H\\
\nonumber
&=&
\varphi_A[k]\theta_A+\varphi_B[k]\theta_B+\varphi_C[k]\theta_C+\varphi_D[k]\theta_D+\varphi_E[k]\theta_E \\
\nonumber
&  &+\varphi_F[k]\theta_F+\varphi_G[k]\theta_G+\varphi_H[k]\theta_H \\
&=& \varphi_1\theta_1
\end{eqnarray}
where
\begin{eqnarray*}
\varphi_1&:=&
\begin{bmatrix}
\varphi_A[k] & \varphi_B[k] & \varphi_C[k] & \varphi_D[k] & \varphi_E[k] & \varphi_F[k] &
\varphi_G[k] & \varphi_H[k]
\end{bmatrix},\\
\theta_1&:=&
\begin{bmatrix}
\theta_A^T & \theta_B^T & \theta_C & \theta_D^T & \theta_E & \theta_F & \theta_G^T & \theta_H^T
\end{bmatrix}^T,
\end{eqnarray*}
with
\begin{eqnarray*}
\varphi_A[k]&:=& CA_0^k\in\mathds{R}^{1\times n_x},  \\
\varphi_B[k]&:=&i_{Fbat}^T[k]\in\mathds{R}^{1\times n_x},\\
\varphi_C[k]&:=&-i_{bat}\in\mathds{R},\\
\varphi_D[k]&:=&v_{Fbat}^T[k]\in\mathds{R}^{1\times n_x},\\
\varphi_E[k]&:=&\boldsymbol{1}[k]\in\mathds{R},\\
\varphi_F[k]&:=&-q_{out}[k]\in\mathds{R},\\
\varphi_G[k]&:=&-\boldsymbol{1}_F^T[k]\in\mathds{R}^{1\times n_x},\\
\varphi_H[k]&:=&q_{outF}^T[k]\in\mathds{R}^{1\times n_x} .
\end{eqnarray*}
Given
\begin{eqnarray}
\label{ThYi}
Y_i&:=&
\begin{bmatrix}
y[1] & \cdots & y[N_{i}] 
\end{bmatrix}^T,\\
\Phi_{i}&:=&
\begin{bmatrix}
 \varphi_{i}[1] \\ \vdots \\ \varphi_{i} [N_{i}]
\end{bmatrix}
=
\begin{bmatrix}
\Phi_A & \Phi_B & \Phi_C & \Phi_D &\Phi_E &\Phi_F & \Phi_G & \Phi_H
\end{bmatrix},
\end{eqnarray}
with $i=1$ and the observer $A_0.$  The parameters can be found by minimising the cost function
\begin{equation}
\label{ThJi}
J_i\left(\theta \right)=\dfrac{1}{2}\left(Y_i-\Phi_i\theta\right)^T\left(Y_i-\Phi_i\theta\right).   
\end{equation}
Taking into account  restrictions \eqref{R1} and \eqref{R2},  the  unknown parameter vector  becomes: 
\begin{equation}
\Theta_1:=
\begin{bmatrix}
\theta_A^T & \theta_B^T & \theta_C & \theta_D^T & \theta_E & \theta_F
\end{bmatrix}^T
\end{equation}
and 
\begin{equation}
 \Phi_1 \theta_1=
\begin{bmatrix}
\Phi_A & \Phi_B & \Phi_C & \Phi_D+\Phi_G\theta_E+\Phi_H\theta_F & \Phi_E & \Phi_F
\end{bmatrix}
 \Theta_1=
\Psi_1 \Theta_1.\\
\end{equation}
Hence \begin{equation}
\label{ThVi}
J_i(\theta)=V_i(\Theta)=\dfrac{1}{2}\left(Y_i-\Psi_i\Theta_i\right)^T\left(Y_i-\Psi_i\Theta_i\right)
\end{equation}
and its minimum occurs when its gradient is zero:
\begin{equation}
\label{gradient}
\left[\dfrac{dV(\Theta_i)}{d\Theta_i}\right]^T=
-\left(\Psi_i+\dfrac{d\Psi_i}{d\Theta_i}\Theta_i\right)^T\left(Y_i-\Psi_i\Theta_i\right)=0
\end{equation}
where
\begin{equation}
\begin{array}{c}
\Psi_1+\dfrac{d\Psi_1}{d\Theta}_1\Theta_1=\\
=\begin{bmatrix}
\Phi_A & \Phi_B & \Phi_C & \Phi_D+\Phi_G\theta_E+\Phi_H\theta_F &  \Phi_E+\Phi_G\theta_D & \Phi_F+\Phi_H\theta_D
\end{bmatrix}
\end{array}
\end{equation}
The minimum of $V_i(\Theta)$ can be found by a Jacobi method where each iteration is given by
\begin{equation}
\label{iteration}
\Theta_i^{(j)}=\left.\left[\left(\Psi_i+\dfrac{d\Psi_i}{d\Theta_i}\Theta_i\right)^T\Psi_i\right]^{-1}\left(\Psi_i+\dfrac{d\Psi_i}{d\Theta_i}\Theta_i\right)\right|_{\Theta=\Theta_i^{(j-1)}}\hspace{-13mm}Y_i.
\end{equation}
The algorithm is initalised with
\begin{equation}
\Theta_i^{(0)}=
\begin{bmatrix}
I_{3n_x+3} & 0_{(3n_x+3)\times 2n_x}
\end{bmatrix}
\hat{\theta}_i,
\end{equation}
where $\hat{\theta}$ is the least squares estimate
\begin{equation}
\label{TheLSqE}
\hat{\theta}_i=\left(\Phi_i^T\Phi_i\right)^{-1}\Phi_i^TY_i.
\end{equation}
\subsubsection{Remaining segments: Known initial OCV and initial state} 
\label{RSegments}
When both $OCV[0]$ and $x[0]$ are known, equation \eqref{eq97} may be written as
\begin{eqnarray}
\nonumber
v_{bat}[k]-OCV_0[k]-CA_0^kx[0]&=&i_{Fbat}^T[k]B_a-i_{bat}[k]R_0+v_{Fbat}^T[k]L-\\
\nonumber
& &-q_{out}[k]\dfrac{1}{C_0}-OCV_{0F}^T[k]L+q_{outF}[k]\dfrac{L}{C_0}.\\
& &
\end{eqnarray}
where
\begin{eqnarray}
OCV_0[k]&=&\boldsymbol{1}[k]OCV[0]\\
OCV_{0F}[k]&=&\boldsymbol{1}_{F}[k]OCV[0]=\left(qI-A_0^T\right)^{-1}C^T\boldsymbol{1}[k]OCV[0].
\end{eqnarray}
Defining
\begin{equation}
y[k]:=v_{bat}[k]-OCV_0[k]-CA_0^kx[0],
\end{equation}
and
\begin{equation}
\varphi_I:=v_{Fbat}[k]-OCV_{0F}[k]
\end{equation}
then
\begin{equation}
y[k]=\varphi_i[k]\theta_i,
\end{equation}
with
\begin{eqnarray}
\varphi_i[k]&:=&
\begin{bmatrix}
\varphi_B[k] & \varphi_C[k] & \varphi_I[k] & \varphi_F[k] & \varphi_H[k]
\end{bmatrix}\\
\theta_i&:=&
\begin{bmatrix}
\theta_B^T & \theta_C & \theta_D^T & \theta_F & \theta_H^T
\end{bmatrix}^T.
\end{eqnarray}
Given \eqref{ThYi},  the observer matrix $A_0$ and
\begin{eqnarray}
\Phi_{i}&=&
\begin{bmatrix}
\varphi_i[1] \\ \vdots \\ \varphi_i[N_{i}]
\end{bmatrix}=
\begin{bmatrix}
\Phi_B & \Phi_C & \Phi_I & \Phi_F & \Phi_H
\end{bmatrix},
\end{eqnarray}
the parameters may be found by minimising \eqref{ThYi} with $i\neq 1$,  taking into consideration restriction \eqref{R2}. Due to this restriction, the true unknown parameter vector is also redefined: 
\begin{equation}
\Theta_i:=
\begin{bmatrix}
\theta_B^T & \theta_C & \theta_D^T & \theta_F
\end{bmatrix}
\end{equation}
and
\begin{equation}
\Phi_i\theta_i=
\begin{bmatrix}
\Phi_B & \Phi_C & \Phi_I+\Phi_H\theta_F & \Phi_F
\end{bmatrix}
\Theta_i=\Psi_i\Theta_i.\\
\end{equation}
As in the previous section,  $\Theta_i$ is estimated by  minimising $V_i(\Theta_i)$  in \eqref{ThVi}  using the Jacobi method  in 	\eqref{iteration} where
\begin{equation}
\Psi_i+\dfrac{d\Psi_i}{d\Theta_i}\theta_i=
\begin{bmatrix}
\Phi_B & \Phi_C & \Phi_I+\Phi_H\theta_F & \theta_F+\Phi_H\theta_D
\end{bmatrix}
\end{equation}
and considering the  initial value:
\begin{equation}
\Theta_i^{(0)}=
\begin{bmatrix}
I_{2n_x+2} & 0_{(2n_x+2)\times n_x}
\end{bmatrix}
\hat{\theta}_i,
\end{equation}
where $\hat{\theta}_i$ given by \eqref{TheLSqE}.

\subsubsection{Observer matrix}
The estimators of Sections \ref{Segment1} and \ref{RSegments} assume  the observer matrix $A_0$ to be  known. Therefore, it must be defined before running the estimators. This matrix is in the companion form of the observable canonical realisation and is only subject to the stability constraint. However, it determines the accuracy of the estimator. Hence,  a natural choice is to choose $A_0$ such that  $V_i(\Theta_i)$   in \eqref{ThVi} is minimised. This can be done using search or gradient methods. As  in this work  we only need to estimate models of order 1 and 2, we chose to use an intensive search method that consists in testing all matrices $A_0$ whose eigenvalues are in a predefined grid, and then choosing the one that leads to the smallest value of $V_i(\Theta_i)$.  To preserve the physical meaning of the estimated models, all eigenvalues of  the search grid  have a positive real part since the cell's discrete-time models result from the sampling of continuous-time models.
\section{CASE STUDY}\label{caseStudy}
The data has been obtained using  
 the discharge board described in \cite{c20} controlled by  an Arduino UNO platform. An  INR18650 F1L Li-ion cell, with capacity of $3350mAh$, standard discharge constant current of 0.2C (650mA) and maximum discharge current of 1.5C (4875mA), was discharged with current pulses of 10 seconds with an amplitude of about 750mA. The time-interval between pulses was also   10 seconds. Both, the battery current and  voltage, where measured with a sampling period of 8ms. 

The measurements were disturbed by noise. A signal to noise ratio (SNR) around 24dB was measured in the voltage measurements, 
 whereas the current noise was filtered with the Kalman filter with \textit{smooth} covariance reset described in \cite{c20}. 
 The voltage was also filtered with the same filter to compute the Best Fit Rate (BFR) index  \eqref{BFR} and to perform the Monte Carlo simulations described later in this section.

\begin{figure}[h!]
\begin{center}
\includegraphics[scale=0.65]{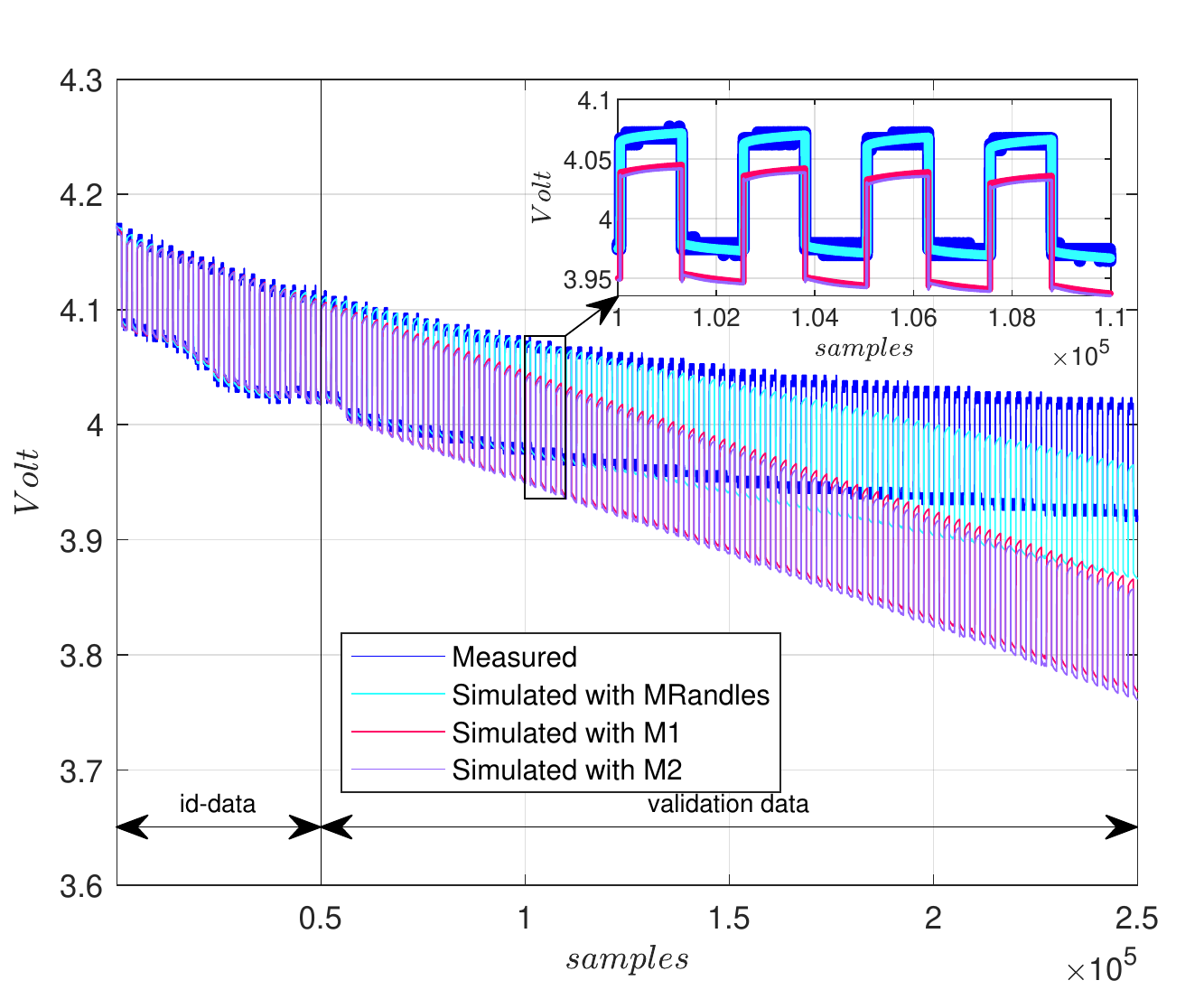}
\caption{ Measured and simulated voltage of the first 250000 points}
\label{Fig12}
\end{center}
\end{figure}

 The first $50000$ data points, corresponding to a time window of $6$ minutes and $40$ seconds, were used to estimate the parameters of the simplified Randles circuit  with the LSE \eqref{SimpleRandleLSQ} in Subsection \ref{simplifiedRC}. In addition, two Thévenin models of orders 1 and 2, respectively, were estimated using the algorithms described in \ref{ModifiedTheveninDm}. In what follows, we  denote by MRandles, M1 and M2 the estimated Randles model and Thévenin models of order 1 and  2, respectively. 
The  Goodness of fit of the models is evaluated by the BFR index:
\begin{equation}
\label{BFR}
BFR(W)=\left(
1-\dfrac{\left\|v_{bat}^0(W)-\hat{v}_{bat}(W)\right\|}{\left\|v_{bat}^0(W)-\bar{v}_{bat}^0(W)\right\|}
\right)\times 100\%,
\end{equation}
where $W$ is the time window, $v_{bat}^0(W)$ the filtered  voltage at the cell terminals in the time window $W$,  $\bar{v}_{bat}^0(W)$ its mean value and $\hat{v}_{bat}(W)$ the simulated voltage.  Table \ref{Table1},  displays the estimated parameters (in IS units),  

\begin{table}[h!]
\begin{center}
\caption{Estimated parameters}
\label{Table1}
\begin{tabular}{l|ccc}
\hline
Parameter & MRandles & M1 & M2 \\
\hline
$OCV(0)$ (Volt) & $4.166$ & $4.165$ & $4.1633$\\
$C_0$ (F) &  $4093.8$ & $2439.3$ & $2368.3$ \\
$A_w$ & $0.0047$ & $-$  & $-$ \\
$R_1$ $(\Omega)$ & $-$ & $0.0153$ &  $0.0183$ \\
$C_1$ (F) & $-$  & $531.69$ &  $211.17$ \\
$R_2$ $(\Omega)$& $-$  & $-$  & $0.0063$ \\
$C_2$ (F) & $-$  &$-$ & $5.7168$\\
$R_b$ ($\Omega$) & $0.1205$ & $0.1206$ & $0.1202$\\
\hline
\end{tabular} 
\end{center}
\end{table}

while Figure~\ref{Fig12} compares the first $250000$ measured values of the battery voltage,  corresponding to a time interval of  33  minutes and 20 seconds, with the voltages simulated  by the estimated models within the window $100000:110000,$ and  in  the validation data set zoomed out.

 This figure clearly shows that the MRandles model is accurate far beyond the window  of the estimation  data. The same is not true for the Thévenin models,  whose accuracy is restricted to this window and   its neighbourhood. This is confirmed by 
  Table \ref{Table1A}  depicts the BFRs calculated in the time windows W0$\equiv 1:50000$ (estimation~data)  W1$\equiv 50001:100000,$  
W2$ \equiv 100001:150000,$ W3$ \equiv 150001:200000$ and W4$\equiv 200001:250000,$ every with $50000$ points ($6$ minutes and $40$ seconds). 
\begin{table}[h!]
\begin{center}
\caption{BFR indexes  ---  First 250000 points of the discharge cycle}
\label{Table1A}
\begin{tabular}{l|ccccc}
\hline
Model & W0 & W1 & W2 & W3 & W4 \\
\hline
MRandles & 94.51\% & 93.06\% & 86.10\% &  54.39\% &    7.24\%\\
M1       & 93.56\% & 68.90\% &  4.50\% & -83.91\% & -184.53\%\\
M2       & 93.06\% &65.74\% & -2.75\% & -95.30\% & -199.81\%\\
\hline
\end{tabular}
\end{center}
\end{table}
 Also   from this table, one can see that the three models exhibit a similar accuracy in the identification window (W0), although the MRrandles model is slightly better. This model is also much more robust to  SOC variation as it keeps good accuracy in the two windows subsequent to the identification data, while the accuracy of the M1 and M2 degrades significantly. To test the robustness of the proposed algorithm to noise, three Monte Carlo simulations were performed. Each simulation  consists of $100$ identification experiments with the filtered current as input and the filtered voltage disturbed by Gaussian white noise as output. The  SNR was defined as the ratio between the filtered voltage peak to peak value and the noise standard deviation. It was of 20dB for the first simulation, 10dB for the second and 0dB for the third. Table \ref{Table2B} shows the mean values of the estimated model parameters with the respective standard deviations (in  brackets), 
\begin{table}
\begin{center}
\caption{Estimated parameters with different noise levels}
\label{Table2B}
\begin{tabular}{c|ccccc}
\hline
SNR (dB) & $OCV(0)$(Volt) & $A_w$ &$C_0$(F)& $R_b(\Omega)$\\
\hline\hline
$20$ & 4.1625  & 3987.3   & 0.0041 & 0.1177\\
     &(0.0005) & (49.6735)&(0.0001)&(0.0001)\\
\hline  
$10$ & 4.1625  & 4009.4  & 0.0041 & 0.1176 \\
     & (0.0015)& (145.11)&(0.0003)& 0.0004)\\
\hline
$0$  & 4.1632  & 4049.0  & 0.0042 & 0.1176\\
 & (0.0052) & (491.30) & (0.0010) &  (0.0014)\\
\hline
\end{tabular}
\end{center}
\end{table}
 while Table \ref{Table3A} depicts the average BFRs   and respective standard deviations (also in brackets). We can observe that even in extreme noise conditions (SNR=0dB) both the parameters and the BFRs in windows W0 and W1 did not suffer significant variations, denoting a high degree of noise immunity of the algorithm 
\begin{table}[h!]
\begin{center}
\caption{BFR of the MRandles model for several noise levels}
\label{Table3A}
\begin{tabular}{c|ccccc}
\hline\hline
SNR (Db) &  W0 & W1 & W2 & W3 & W4 \\
\hline
 20 & 95.13\%  & 94.30\% & 83.94\% & 49.31\% & -0.04\% \\
      & (0.003\%)& (0.14\%)& (1.47\%)& (2.66\%)& (3.69\%)\\
 \hline
 10 & 95.07\% & 94.00\% & 84.24\% & 50.18\% & 1.21\%\\
      & (0.02\%)&(0.54\%)&(4.04\% )& (7.60\%)& 10.60\%\\
 \hline
 0  & 94.60\% &  91.96 \% & 80.10\% &  48.78\% & 0.23\%\\
      &(0.22\%) & (2.60\%) & (9.32\%) & (22.93\%) & (33.36\%)\\
\hline
\end{tabular}
\end{center}
\end{table}

 Figure \ref{Fig12} and Table \ref{Table1A} show that an entire discharge cycle cannot be described by an LTI model. This is because the model parameters depend on  SOC. To determine this dependence the discharge cycle was split into several segments, and an LTI Randles model was identified for each segment. In this way, we obtained an algorithm that identifies a piecewise LTI model capable of describing the entire discharge cycle. Moreover, since the OCV is a state of the model, it can be estimated by a suitable observer and its dependence on the  SOC can be determined without performing a time consuming experiment. 
\begin{figure}[h!]
\begin{center}
\includegraphics[scale=0.6]{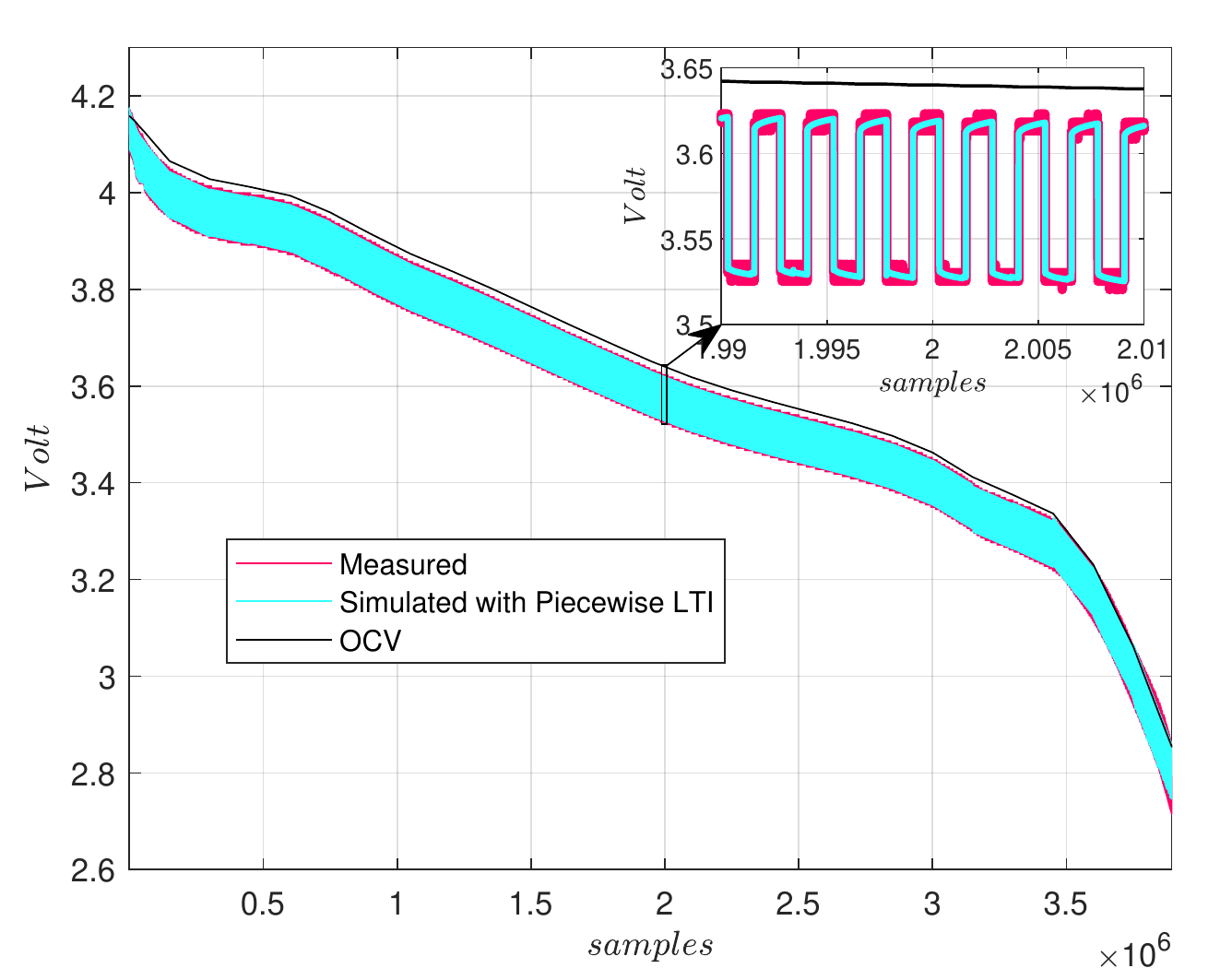}
\caption{Discharge in the entire  SOC}
\label{Fig13}
\end{center}
\end{figure}
 Figure \ref{Fig13} compares the measured battery voltage with the simulated by a piecewise LTI model with segments of 150000 data points identified by the algorithm. It also depicts the OCV estimated by the model. The overal BFR was $93.77\%$ denoting    the model   high  accuracy.

\end{document}

%% file: OCV.tex
\begin{circuitikz}
\draw(0,0)
to[short,-o](1.5,0);
\draw(0,0)
to[C=$C_0(t)$] (0,2) 
to[short,-o](1.5,2);
\draw(1.5,1.9)
to[open,v^>=$OCV(t)$](1.5,0);
\draw(1.5,2)node[below]{$+$};
\draw(1.5,-0.1)node[above]{$-$};
\draw[->](0.3,2.2)--(1.3,2.2);
\draw(0.8,2.2)node[above]{$i_{bat}(t)$};
\end{circuitikz}

%% file: Rint.tex
\begin{circuitikz}
\draw(0,0)
to[short,-o](1.5,0)
to[short,-o](4,0);

\draw(0,0)
to[C=$C_0(t)$] (0,2) 
to[short,-o](1.5,2)
to[short](1.8,2)
to[R=$R_0$](3,2) 
to[short,-o](4,2);

\draw(1.5,1.9)
to[open,v^>=$OCV(t)$](1.5,0);
\draw(1.5,2)node[below]{$+$};
\draw(1.5,-0.1)node[above]{$-$};

\draw(4,1.9)
to[open,v^>=$v_{bat}(t)$](4,0);
\draw(4,2)node[below]{$+$};
\draw(4,-0.1)node[above]{$-$};

\draw[->](3,2.2)--(4,2.2);
\draw(3.5,2.2)node[above]{$i_{bat}(t)$};
\end{circuitikz}

%% file: randlesCircuit.tex
\begin{circuitikz}
\draw(0,0)
to[C=$C_0$] (0,2) 
to[short,-o](0.7,2)
to[short](1,2)
to[R=$R_0$](3,2) 
to[short](3.5,2);

\draw(3.5,2)
to[short](3.5,2.9)
to[C=$C_{dl}$](8,2.9) 
to[short](8,2);

\draw(3.5,2.6)
to[open,v=$v_{dl}$](8,2.6);
\draw(3.8,2.9) node[below]{$+$};
\draw(7.7,2.9) node[below]{$-$};

\draw [->] (6.5,3.1)--(7.3,3.1);
\draw (6.9,3.1) node [above]{$i_{dl}$};

\draw(3.5,2)
to[short](3.5,1.1)
to[generic=$Z_w$](5.5,1.1)
to[short](6,1.1)
to[R=$R_{ct}$](8,1.1)
to[short](8,2);

\draw(3.5,1.5)
to[open,v^>=$v_w$](5.5,1.5);

\draw[->] (5.6,0.9)--(6.4,0.9);
\draw(6,0.9) node[below]{$i_{w}$};
\draw(3.7,1.1) node[above]{$+$};
\draw(5.3,1.1) node[above]{$-$};
\draw(8,2)
to[short,-o](9.5,2);
\draw(0,0)
to[short](0.7,0)
to[short,o-o](9.5,0);
\draw(9.5,2)
to[open,v^>=$v_{bat}$](9.5,0);
\draw(0.7,2)
to[open,v^>=$OCV$](0.7,0);
\draw(0.7,2)node[below]{$+$};
\draw(0.7,0)node[above]{$-$};
\draw(9.5,2) node[right]{$+$};
\draw(9.5,0) node[right]{$-$};
\draw[->](8.2,2.2)--(9.2,2.2);
\draw(8.7,2.2)node[above]{$i_{bat}$};
\end{circuitikz}

%% file: ZwBodeDiagrams.tex
\begin{tikzpicture}[scale=0.75]
\draw [->] (0,0)--(12,0) node [below] {$\omega$};
\draw [->] (0,-2)--(0,0) node[left]{$0$} -- (0,4) node [right] {$dB$};
\draw(0.05,0) node [below] {$0.1A_w^2$};
\draw[blue,thick] (0,3)node[black,left] {$10$}--(9,0) node[black, below]{$A_w^2$}--(12,-1);
\draw[->] (0,-5)--(0,-2) node[left]{Degrees};
\draw[->] (0,-2.5)node[left]{$0^o$}--(12,-2.5) node [below] {$\omega$};
\draw[blue,thick] (0,-4) node[black,left]{$-45^o$}--(12,-4);

\end{tikzpicture} 

%% file: warburgImpulse.tex
\begin{tikzpicture}[xscale=0.8, yscale=1]

   \draw[->] (-0.013,0) -- (13.39,0) node [above] {$k$};
   \draw[->] (0,0) -- (0,3) node [left]{$\dfrac{w[k]}{A_w\sqrt{T_s}}$};

   \input{walburgImpulse-Table}
\end{tikzpicture}

%% file: compareWarburg.tex
\begin{tikzpicture}[xscale=0.8, yscale=1]

   \draw[->] (-0.013,0) -- (13.39,0) node [above] {$k$};
   \draw[->] (0,0) -- (0,3) node [left]{$\dfrac{w[k]}{A_w\sqrt{T_s}}$};

   \input{walburgImpulse-Table}
   \input{walApproxImpulse-Table}
\end{tikzpicture}

%% file: bodePlots.tex
\begin{tikzpicture}[xscale=3.6, yscale=0.1]

   \draw (-3,-5) --(0,-5)--(0,30)--(-3,30)--(-3,-5);
   \draw(-2.6,30) node[above]{$20\log\left|\dfrac{Z_w}{A_w\sqrt{T_s}}\right|$ (dB)};
   
   \draw (-3,-5) node[left] {$-5$};   
   \draw[dashed] (-3,0) node[left]{$0$}--(0,0);
   \draw[dashed] (-3,5) node[left]{$5$}--(0,5);
   \draw[dashed] (-3,10) node[left]{$10$}--(0,10);
   \draw[dashed] (-3,15) node[left]{$15$}--(0,15);
   \draw[dashed] (-3,20) node[left]{$20$}--(0,20);
   \draw[dashed] (-3,25) node[left]{$25$}--(0,25);
   \draw (-3,30) node[left] {$30$}; 

   \draw [dashed] (-2.6690,-5)--(-2.6690,30); 
   \draw [dashed] (-2.5229,-5)--(-2.5229,30); 
   \draw [dashed] (-2.3979,-5)--(-2.3979,30); 
   \draw [dashed] (-2.3010,-5)--(-2.3010,30); 
   \draw [dashed] (-2.2218,-5)--(-2.2218,30); 
   \draw [dashed] (-2.1549,-5)--(-2.1549,30); 
   \draw [dashed] (-2.0969,-5)--(-2.0969,30); 
   \draw [dashed] (-2.0458,-5)--(-2.0458,30); 
   
   \draw [dashed] (-1.6690,-5)--(-1.6690,30); 
   \draw [dashed] (-1.5229,-5)--(-1.5229,30); 
   \draw [dashed] (-1.3979,-5)--(-1.3979,30); 
   \draw [dashed] (-1.3010,-5)--(-1.3010,30); 
   \draw [dashed] (-1.2218,-5)--(-1.2218,30); 
   \draw [dashed] (-1.1549,-5)--(-1.1549,30); 
   \draw [dashed] (-1.0969,-5)--(-1.0969,30); 
   \draw [dashed] (-1.0458,-5)--(-1.0458,30); 
   
   \draw[dashed] (-1,-5) node[below] {$0.1$}--(-1,30);
   \draw [dashed] (-0.6690,-5)--(-0.6690,30); 
   \draw [dashed] (-0.5229,-5)--(-0.5229,30); 
   \draw [dashed] (-0.3979,-5)--(-0.3979,30); 
   \draw [dashed] (-0.3010,-5)--(-0.3010,30); 
   \draw [dashed] (-0.2218,-5)--(-0.2218,30); 
   \draw [dashed] (-0.1549,-5)--(-0.1549,30); 
   \draw [dashed] (-0.0969,-5)--(-0.0969,30); 
   \draw [dashed] (-0.0458,-5)--(-0.0458,30); 

   \input{bodeMag-Table} 
   \draw [blue,thick] (-3,25.029)--(0, -4.9715); 
   
   \draw (-3,-45) --(0,-45)--(0,-10)--(-3,-10)--(-3,-45);
   \draw(-2.7,-10) node [above]{$\angle Z_w$ (degrees)};
   
   \draw(-3,-45) node[left]{$-180$};
   \draw[dashed] (-3,-40) node[left]{$-160$}--(0,-40);
   \draw[dashed] (-3,-35) node[left]{$-140$}--(0,-35);
   \draw[dashed] (-3,-30) node[left]{$-120$}--(0,-30);
   \draw[dashed] (-3,-25) node[left]{$-100$}--(0,-25);
   \draw[dashed] (-3,-20) node[left]{$-80$}--(0,-20);
   \draw[dashed] (-3,-15) node[left]{$-60$}--(0,-15);
   \draw(-3,-11.25) node[left]{$-45$};
 
   \draw(-3,-45) node [below] {$0.001$};
   \draw [dashed] (-2.6690,-45)--(-2.6690,-10); 
   \draw [dashed] (-2.5229,-45)--(-2.5229,-10); 
   \draw [dashed] (-2.3979,-45)--(-2.3979,-10); 
   \draw [dashed] (-2.3010,-45)--(-2.3010,-10); 
   \draw [dashed] (-2.2218,-45)--(-2.2218,-10); 
   \draw [dashed] (-2.1549,-45)--(-2.1549,-10); 
   \draw [dashed] (-2.0969,-45)--(-2.0969,-10); 
   \draw [dashed] (-2.0458,-45)--(-2.0458,-10); 
   
   \draw[dashed] (-2,-45) node[below] {$0.01$}--(-2,-10);
   \draw [dashed] (-1.6690,-45)--(-1.6690,-10); 
   \draw [dashed] (-1.5229,-45)--(-1.5229,-10); 
   \draw [dashed] (-1.3979,-45)--(-1.3979,-10); 
   \draw [dashed] (-1.3010,-45)--(-1.3010,-10); 
   \draw [dashed] (-1.2218,-45)--(-1.2218,-10); 
   \draw [dashed] (-1.1549,-45)--(-1.1549,-10); 
   \draw [dashed] (-1.0969,-45)--(-1.0969,-10); 
   \draw [dashed] (-1.0458,-45)--(-1.0458,-10); 
   
   \draw[dashed] (-1,-45) node[below] {$0.1$}--(-1,30);
   \draw [dashed] (-0.6690,-45)--(-0.6690,-10); 
   \draw [dashed] (-0.5229,-45)--(-0.5229,-10); 
   \draw [dashed] (-0.3979,-45)--(-0.3979,-10); 
   \draw [dashed] (-0.3010,-45)--(-0.3010,-10); 
   \draw [dashed] (-0.2218,-45)--(-0.2218,-10); 
   \draw [dashed] (-0.1549,-45)--(-0.1549,-10); 
   \draw [dashed] (-0.0969,-45)--(-0.0969,-10); 
   \draw [dashed] (-0.0458,-45)--(-0.0458,-10); 
   
   \draw(0,-45) node [below] {$1$};
   \draw(-1.5229,-47) node[below]{$\omega/\omega_N$};

   \input{bodePhase-Table} 
   \draw[blue,thick] (-3,-11.25)--(0,-11.25);
\end{tikzpicture}

%% file: bodePlotsC.tex
\begin{tikzpicture}[xscale=1.4, yscale=0.035]

   \draw (-6,-40) --(2,-40)--(2,60)--(-6,60)--(-6,-40);
   \draw(-5,60) node[above]{$20\log\left|\dfrac{Z_w}{A_w\sqrt{T_s}}\right|$ (dB)};
   
   \draw (-6,-40) node[left] {$-40$};   
   \draw[dashed] (-6,-20) node[left]{$-20$}--(2,-20);
   \draw[dashed] (-6,0) node[left]{$0$}--(2,0);
   \draw[dashed] (-6,20) node[left]{$20$}--(2,20);
   \draw[dashed] (-6,40) node[left]{$40$}--(2,40);
   \draw (-6,60) node[left] {$60$}; 

   \draw [dashed] (-5,-40)--(-5,60);
   \draw [dashed] (-4,-40)--(-4,60);
   \draw [dashed] (-3,-40)--(-3,60);
   \draw [dashed] (-2,-40)--(-2,60);
   \draw [dashed] (-1,-40)--(-1,60);
   \draw [dashed] (0,-40)--(0,60); 
   \draw [dashed] (1,-40)--(1,60);

   \input{bodeMagC-Table} 
   \draw [blue,thick] (-6,59)--(2, -21); 
   
  \draw (-6,-150) --(2,-150)--(2,-55)--(-6,-55)--(-6,-150);
   \draw(-5,-55) node [above]{$\angle Z_w$ (degrees)};
   \draw[dashed] (-6,-140) node[left]{$-80$}--(2,-140);
   \draw[dashed] (-6,-120) node[left]{$-60$}--(2,-120);
   \draw[dashed] (-6,-100) node[left]{$-40$}--(2,-100);
   \draw[dashed] (-6,-80) node[left]{$-20$}--(2,-80);
   \draw[dashed] (-6,-60) node[left]{$0$}--(2,-60);
   \draw(-6,-150) node [below] {$10^{-6}$};
   \draw [dashed] (-5,-150)node [below] {$10^{-5}$}--(-5,-55);
   \draw [dashed] (-4,-150)node [below] {$10^{-4}$}--(-4,-55);
   \draw [dashed] (-3,-150)node [below] {$10^{-3}$}--(-3,-55);
   \draw [dashed] (-2,-150)node [below] {$0.01$}--(-2,-55);
   \draw [dashed] (-1,-150)node [below] {$0.1$}--(-1,-55);
   \draw [dashed] (0,-150)node [below] {$1$}--(0,-55); 
   \draw [dashed] (1,-150)node [below] {$10$}--(1,-55); 
   \draw(2,-150) node [below] {$100$};
%
%
%
  \draw(-2,-160) node[below]{$\omega/T_s$};
%
   \input{bodePhaseC-Table} 
   \draw[blue,thick] (-6,-105)--(2,-105);
   \draw(-6,-107)node[left]{$-45$};
\end{tikzpicture}

%% file: seriesRCParallell.tex
\ctikzset{bipoles/length=.5cm}
\ctikzset{bipole label style/.style={font=\tiny}}
\begin{circuitikz}[xscale=1.5,yscale=1.5]
\draw(0,0)
to[C=$C_0$] (0,1) 
to[short,-o](0.35,1)
to[R=$R_0$,-o](1,1) 
to[short](1.2,1);

\draw(1.2,1)
to[short](1.2,1.25)
to[C=$C_1$](1.75,1.25) 
to[short](1.75,1);
\draw(1.2,1)
to[short](1.2,0.75)
to[R=$R_1$](1.75,0.75)
to[short](1.75,1);

\draw(1.75,1)
to[short](2,1);

\draw(2,1)
to[short](2,1.25)
to[C=$C_2$](2.55,1.25) 
to[short](2.55,1);
\draw(2,1)
to[short](2,0.75)
to[R=$R_2$](2.55,0.75)
to[short](2.55,1);

\draw (2.55,1)--(2.8,1);
\draw [dashed] (2.8,1)--(3.5,1);

\draw(3.6,1)
to[short](3.85,1);

\draw(3.85,1)
to[short](3.85,1.25)
to[C=$C_n$](4.4,1.25) 
to[short](4.4,1);
\draw(3.85,1)
to[short](3.85,0.75)
to[R=$R_n$](4.4,0.75)
to[short](4.4,1);

\draw(4.4,1)
to[R=$R_{ct}$](5.4,1)
to[short,-o](5.5,1);

\draw(1,1)
to[short](1,1.75)
to[C=$C_{ct}$](5.5,1.75)
to[short](5.5,1);

\draw(5.5,1)
to[short,-o](6.5,1);

\draw[->](5.75,1.1)--(6.25,1.1);
\draw(6,1.1)node[above]{\tiny{$i_{bat}$}};

\draw(0,0)
to[short](0.35,0)
to[short,o-o](6.5,0);

\draw(6.5,1)
to[open,v^>=\tiny $v_{bat}$](6.5,0);
\draw(6.5,1) node[right]{\tiny $+$};
\draw(6.5,0) node[right]{\tiny $-$};

\draw(0.4,1)
to[open,v^>=\tiny $OCV$](0.4,0);
\draw(0.3,1)node[below]{\tiny $+$};
\draw(0.3,0)node[above]{\tiny $-$};
\end{circuitikz}

%% file: simplifiedRandlesCircuit.tex
\begin{circuitikz}
\draw(0,0)
to[C=$C_0$] (0,2) 
to[short,-o](0.7,2)
to[short](1,2)
to[R=$R_b\equiv R_0+R_{ct}$](3,2) 
to[short](3.5,2)
to[generic=$Z_w$](5.5,2)
to[short,-o](7,2);

\draw(3.5,2.4)
to[open,v^>=$v_w$](5.5,2.4);

\draw[->] (5.6,2.2)--(6.4,2.2);
\draw(6,2.2) node[above]{$i_{bat}$};
\draw(3.7,2) node[above]{$+$};
\draw(5.3,2) node[above]{$-$};

\draw(7,2)
to[open,v^>=$v_{bat}$](7,0);

\draw(7,2) node[right]{$+$};
\draw(7,0) node[right]{$-$};

\draw(0,0)
to[short](0.7,0)
to[short,o-o](7,0);

\draw(0.7,2)
to[open,v^>=$OCV$](0.7,0);
\draw(0.7,2)node[below]{$+$};
\draw(0.7,0)node[above]{$-$};

\end{circuitikz}

%% file: seriesRCParallell0.tex
\ctikzset{bipoles/length=.5cm}
\ctikzset{bipole label style/.style={font=\tiny}}
\begin{circuitikz}[xscale=2,yscale=2]
\draw(0,0)
to[C=$C_0$] (0,1) 
to[short,-o](0.35,1)
to[R=$R_0$,-o](1,1) 
to[short](1.2,1);

\draw(1.2,1)
to[short](1.2,1.25)
to[C=$C_1$](1.75,1.25) 
to[short](1.75,1);
\draw(1.2,1)
to[short](1.2,0.75)
to[R=$R_1$](1.75,0.75)
to[short](1.75,1);

\draw(1.75,1)
to[short](2,1);

\draw(2,1)
to[short](2,1.25)
to[C=$C_2$](2.55,1.25) 
to[short](2.55,1);
\draw(2,1)
to[short](2,0.75)
to[R=$R_2$](2.55,0.75)
to[short](2.55,1);

\draw (2.55,1)--(2.8,1);
\draw [dashed] (2.8,1)--(3.5,1);

\draw(3.6,1)
to[short](3.85,1);

\draw(3.85,1)
to[short](3.85,1.25)
to[C=$C_{n_x}$](4.4,1.25) 
to[short](4.4,1);
\draw(3.85,1)
to[short](3.85,0.75)
to[R=$R_{n_x}$](4.4,0.75)
to[short](4.4,1);

\draw(4.4,1)
to[short,-o](4.8,1);

\draw[->](4.5,1.1)--(4.75,1.1);
\draw(4.6,1.1)node[above]{\tiny{$i_{bat}$}};

\draw(0,0)
to[short](0.35,0)
to[short,o-o](4.8,0);

\draw(4.8,1)
to[open,v^>=\tiny $v_{bat}$](4.8,0);
\draw(4.8,1) node[right]{\tiny $+$};
\draw(4.8,0) node[right]{\tiny $-$};

\draw(0.4,1)
to[open,v^>=\tiny $OCV$](0.4,0);
\draw(0.3,1)node[below]{\tiny $+$};
\draw(0.3,0)node[above]{\tiny $-$};

\draw(0.5,0.6)
to[open,v_>=\tiny $v_s$](5.1,0.6);
\draw(1.2,0.75)node[below]{\tiny $+$};
\draw(4.4,0.75)node[below]{\tiny $-$};

\end{circuitikz}